\documentclass[onecolumn,11pt,a4paper,aps]{revtex4}

\newcommand{\eq}[1]{\begin{equation}#1\end{equation}}
\newcommand{\dd}{\mathrm{d}}
\newcommand{\ee}{\mathrm{e}}
\newcommand{\ex}{\mathrm{exp}}
\newcommand{\ch}{\mathrm{ch\,}}
\newcommand{\sh}{\mathrm{sh\,}}
\newcommand{\thf}{\mathrm{th\,}}
\usepackage{amsmath}
\usepackage{graphics}
\usepackage{graphicx}
\usepackage{psfrag}

\begin{document}

\title{Solution of the fermionic entanglement problem with interface defects}
\author{Viktor Eisler$^1$ and Ingo Peschel$^2$}
\affiliation{
$^1$Niels Bohr Institute, University of Copenhagen, Blegdamsvej 17,
DK-2100 Copenhagen \O, Denmark \\
$^2$Fachbereich Physik, Freie Universit\"at Berlin,
Arnimallee 14, D-14195 Berlin, Germany
}

\begin{abstract}
We study the ground-state entanglement of two halves of a critical transverse Ising
chain, separated by an interface defect. From the relation to a two-dimensional 
Ising model with a defect line we obtain an exact expression for the continuously
varying effective central charge. The result is relevant also for other fermionic chains.
\end{abstract}

\maketitle
\section{Introduction}

In one-dimensional systems, already a local perturbation can have important
and interesting effects. This holds for the transport properties, but also
for the entanglement between two pieces of a quantum chain. The influence of 
a defect located at the interface between them was first investigated by Levine 
for a Luttinger liquid \cite{Levine04}. In this case, also studied on a lattice 
in \cite{Zhao/Peschel/Wang06}, the interaction leads to a strong renormalization
of the defect strength \cite{Kane/Fisher92} and thus to special properties.
The situation is simpler for a fermionic hopping model, or XX chain, which was 
investigated in \cite{Peschel05}. In this case it was found that the logarithmic 
dependence of the entanglement entropy $S$ on the subsystem size $L$ remains 
intact, but the prefactor depends continuously on the defect strength. Thus, 
for $\nu$ interfaces 
\eq{
S = \nu \,\frac {c_\mathrm{eff}} {6} \ln L + k
\label{ceff}}
with a variable coefficient which replaces the central charge $c$ of the pure 
system. The dependence of $c_\mathrm{eff}$ on the defect was determined rather 
precisely by numerical calculations, both in XX chains and in transverse Ising (TI) 
chains \cite{Igloi/Szatmari/Lin09}. But although a simple analytical expression fits
the data qualitatively, no exact formula has been given so far.

In this paper, we want to look at this problem once more and present an analytical
treatment based on the connection with a two-dimensional classical system with a
defect line. For the TI quantum chain, this is a two-dimensional Ising model where
defect lines are in fact marginal perturbations which lead to variable local 
magnetic exponents and have been studied over the years by various methods, see
\cite{Igloi/Peschel/Turban93} for a review and \cite{Oshikawa/Affleck97} for a 
more recent approach using boundary conformal field theory. Our techniques will be 
simpler and purely fermionic. Using conformal mapping, we will reduce the 
considerations to the study of the transfer matrix in a strip with defect lines 
parallel to the edges \cite{Turban85}. From its single-particle excitations we will 
obtain not only the spectrum of the reduced density matrix, but also the formula 
for the effective central charge. Our approach is similar to a recent calculation 
by Sakai and Satoh, who treated conformal interfaces between two different bosonic 
systems with $c=1$ \cite{Sakai/Satoh08}. Actually, we were motivated by this work, 
because the formula found there does not describe the results found in the fermionic 
XX and TI chains. Nevertheless, our final result will turn out to be closely related.

In the following Section 2 we set up the problem and present some numerical results 
for the spectrum of the reduced density matrix for later comparison. In Section 3 
we describe the connection of the TI chain with the two-dimensional Ising model,
including the necessary renormalization of parameters. In Section 4 we determine
the proper transfer matrix in the strip and its single-particle spectrum. This
allows us to find the entanglement entropy in Section 5 and to give a closed
formula for $c_\mathrm{eff}$. En route, we also obtain the largest eigenvalue
of the reduced density matrix. In Section 6 we sum up our findings and discuss the 
extension to other chains and defects. Additional details of the transfer matrix 
calculation are given in Appendix A, while in Appendix B we show that with our 
spectra one can also rederive the local magnetic exponent, which is a check on the
approach and a further application.

\section{Chain problem}

In the following we will study the transverse Ising chain with $2L$ sites,
open ends and Hamiltonian
\eq{
H=-\frac {1}{2}\sum_{n=-L+1}^{L-1} J_n \sigma_n^z\sigma_{n+1}^z
  -\frac {1}{2}\sum_{n=-L+1}^{L} h_n\sigma_n^x
\label{HamTI}}
where the $\sigma_n^{\alpha}$ are Pauli matrices at site $n$. We are interested in 
the entanglement of one of the half-chains with the other in the ground state. To
have a modified bond at the interface, the parameters are chosen as $h_n=h$, $J_n=J$ 
for $n \not= 0$ and $J_0=t\,J$. Similarly, a modified transverse field at the interface 
is described by $h_n=h$ for $n \not= 0$ and $h_0=t\,h$. We work at the critical point 
given by $J=h=1$. Then $H$ is normalized such that the velocity of the single-particle
excitations in the homogeneous system is $v=1$ and only one parameter, the defect 
strength $t$, remains.

To calculate the entanglement, one introduces fermions in (\ref{HamTI}) via the
Jordan-Wigner transformation and then determines the fermionic single-particle
eigenvalues $\varepsilon_l$ of the operator $\mathcal{H}$ in the exponent
of the reduced density matrix
\eq{ 
\rho = \frac{1}{Z} \exp(-\mathcal{H})
\label{rho}}
from the correlations \cite{Peschel/Eisler09}. The entanglement entropy 
$S= -\mathrm{Tr}\,(\rho \ln \rho)$ is then given by
\eq{
S = \sum_l \ln (1 + \mathrm{e}^{-\varepsilon_l})+\sum_l
\frac{\varepsilon_l}{\mathrm{e}^{\varepsilon_l} + 1}
\label{ent-thermo}}
Numerical calculations along these lines were done in \cite{Igloi/Szatmari/Lin09} for
the case of a ring with $\emph{two}$ equal defects on opposite sides. This configuration
has essentially the same features as the open chain and one obtains the result
(\ref{ceff}) with $\nu=2$. The effective central charge $c_\mathrm{eff}(t)$ rises 
from zero for $t=0$ (where the system is cut in two) to one-half for $t=1$ (where 
it becomes homogeneous). It is shown in Fig. 2 of Ref. \cite{Igloi/Szatmari/Lin09}. 

The same function had been found before, also numerically, for the case of a segment 
in an infinite XX chain with a hopping defect at one interface \cite{Peschel05}. The other,
unmodified interface then contributes an additive constant to $c_\mathrm{eff}(t)$. The 
background of this universality is the close connection between the two kinds of chains. 
By taking a critical TI chain with a bond defect and another one with a field defect, both 
of strength $t$, superimposing them and using dual variables, one arrives at 
\cite{Peschel/Schotte84,Turban84}
\eq{
H=-\frac {1}{2} \sum_{n\not=0} (\sigma_n^x\sigma_{n+1}^x
    +\sigma_n^y\sigma_{n+1}^y)
  -\frac {1}{2}\, t\,(\sigma_0^x\sigma_{1}^x+\sigma_0^y\sigma_{1}^y)
\label{HamXX}}
It was shown in \cite{Igloi/Juhasz08} that the XX system with $2L$ sites in the subsystem
and the combined TI systems with $L$ sites have the same set of eigenvalues $\varepsilon_l$.
This holds even in the inhomogeneous case and can easily be checked numerically.
As a consequence, $S_{XX}(2L)=S^1_{TI}(L)+S^2_{TI}(L)$, which explains the common 
$c_\mathrm{eff}$.

\begin{figure}
\center
\includegraphics[scale=0.7]{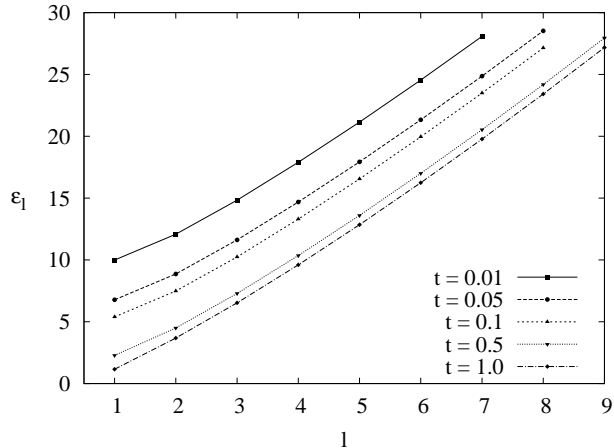}
\caption{Single-particle eigenvalues $\varepsilon_l$ as a function of the defect
strength for TI chains with $2L=300$ sites.}
\label{fig:spectra}
\end{figure}

Given the central role of the single-particle spectrum of $\mathcal{H}$, it is
important to see how this spectrum varies with the defect strength. This is shown
in Fig. \ref{fig:spectra} for TI chains with $2L=300$ sites. 
%
%
%
%
The characteristic feature is the development of a gap at the lower end of
the spectrum and an upward shift of the whole dispersion curve, which increases
roughly like $\ln(1/t)$ as $t$ goes to zero. This upward shift causes a decrease
of $S$ for a fixed value of $L$. In particular, it makes $S$ vanish if $t=0$ and
the system is cut in two. The logarithmic variation of $S$ with $L$, on the other
hand, is related to a gradual lowering of the eigenvalues as $L$ increases. These
features will be found again in the analytical treatment, to which we now turn. 

\section{Relation to a two-dimensional problem}

\begin{figure}
\center
\includegraphics[scale=0.5]{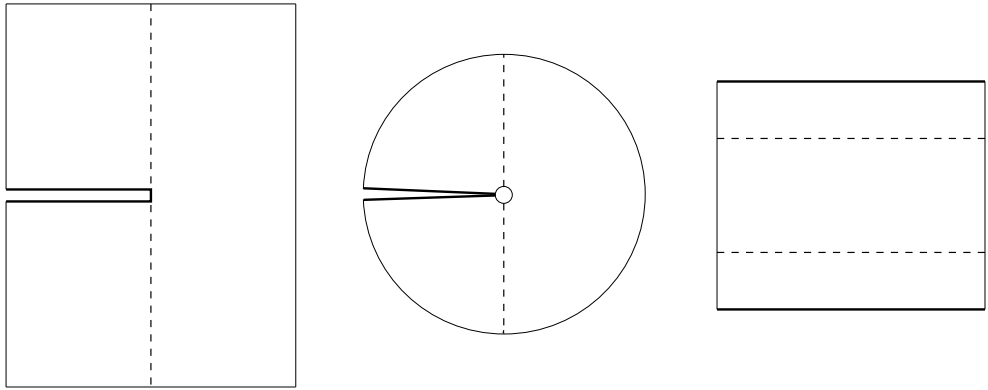}
\caption{Representation of $\rho$ for a half-chain by two-dimensional partition functions. 
Left: Original representation. Centre: Simplified annular geometry. Right: Strip geometry 
obtained by mapping. The defect line is shown dotted.}
\label{fig:confgeom}
\end{figure}

In the case of a homogeneous chain, the TI Hamiltonian commutes with the diagonal
transfer matrix of an isotropic two-dimensional Ising model and its ground state
can be obtained from the partition function of a long Ising strip 
\cite{Peschel/Kaulke/Legeza99,Peschel/Eisler09}. The reduced density matrix is
then given by a strip with a perpendicular cut as shown in Fig. \ref{fig:confgeom}
on the left. One expects that the chain with a defect is related to a planar Ising system
with a defect line as indicated. However, the relations in \cite{Igloi/Lajko96} show that 
by diagonally layering the Ising lattice one cannot obtain a commuting TI Hamiltonian 
with only a $\emph{single}$ modification (bond or field). Therefore we consider a 
lattice in the normal orientation and its row or column transfer matrices. It is 
well-known, that in a proper anisotropic (Hamiltonian) limit, they are just the 
exponential of a TI operator. This also holds with defect lines. For the case of a 
ladder defect, we can refer to Fig. \ref{fig:latgeom}. If the horizontal couplings are
large and the vertical ones are small, the column transfer matrix, running in the direction 
of the ladder, is given by
\eq{
T=A \,\exp(-2K_2^*\,H)
\label{TransferH}}
with $H$ as in (\ref{HamTI}) and $A$ a constant. The quantity $K_2^*$ is the dual coupling 
of $K_2$, $\sh(2K_2^*)=1/\sh(2K_2)$, and thus also small by assumption. The parameters in 
$H$ are $h_n=1$, $J_n=K_1/K_2^*$, $J_0=K_0/K_1\,J_n$. At the critical point $K_2^*=K_1$ and
one obtains the critical TI model with a bond defect given by $t=K_0/K_1$.

%
%
%
%
This is, however, not quite enough, since we want to use a conformal mapping in the
following, for which the 2D system should be isotropic. In the homogeneous case, one
can achieve this by a rescaling such that the velocity of the excitations becomes
$v=1$. Thus one omits the factor $2K_2^*$ in (\ref{TransferH}) and has $T=A \exp(-H)$.
This is then valid in the continuum limit or for large distances. With a defect, the
situation is more subtle. The local magnetic exponent $\beta_1$, which also gives
the power-law decay of the correlations parallel to the defect, is for the
ladder \cite{Bariev79,McCoy/Perk80,Igloi/Peschel/Turban93} 
\eq{
\beta_1 = \frac {2}{\pi^2} \arctan^2 \left(\frac{\thf K_1}{\thf K_0}\right)
\label{betaladder}}
where the couplings are taken at criticality. Thus it matters whether one is in the
Hamiltonian limit, where the argument in the $\arctan$ is $K_1/K_0$, or in the isotropic
case, where it is $\thf K_1/\thf K_0$ with 
$\thf K_1=\thf K_c=\sqrt 2 -1$. Therefore the TI 
Hamiltonian only corresponds to an isotropic system if also the defect parameter $t$ 
is renormalized to
\eq{
t = \frac {\thf K_0}{\thf K_1} ,\quad \mbox{bond defect}
\label{tbond}}
For a chain defect with modified horizontal couplings $K_0$, analogous considerations
apply. The TI Hamiltonian then contains a modified field given by $K_0^*/K_2^*$ where
now the dual couplings enter. These also appear in the local exponent and one has to
renormalize the defect parameter to   
\eq{
t = \frac {\thf K_0^*}{\thf K_2^*} ,\quad  \mbox{field defect}  
\label{tfield}}
where one can use $K_2^*=K_1=K_c$.

Having related the TI chain to an isotropic Ising lattice, we can now turn to the 
partition function. Instead of studying the geometry shown on the left in Fig. \ref{fig:confgeom},
we will consider the one in the centre, where the system is an annulus with a small inner
radius $a$ of the order of the lattice constant and outer radius $R=La$. This could be 
achieved approximately by the mapping used in \cite{Holzhey94}. The pieces in the lower
and upper half-plane can now be viewed as azimuthal corner transfer matrices (CTMs) 
spanning an angle of $180^\circ$ and containing a defect line in their centre. Using the 
mapping $w=\ln z$, the annulus becomes the strip shown on the right of Fig. \ref{fig:confgeom}
with width $\ln(R/a)$ and height $2\pi$. One can now calculate the partition function $Z(n)$ 
for a strip formed by repeating $n$ such units to give a height of $2\pi n$, and closed 
in the vertical direction. Then $S$ follows from the usual formula, used also in 
\cite{Sakai/Satoh08}, 
\eq{
S= \left. \left( 1-\frac {\dd}{\dd n} \right) \ln Z(n) \right|_{n=1}.
\label{entropydiff}}
However, since the transfer matrix needed to calculate $Z(n)$ is just the transform
of the corner transfer matrix which upon squaring gives $\rho$, one can also obtain 
the single-particle eigenvalues in $\mathcal{H}$ directly from it and then use 
(\ref{ent-thermo}).

\section{Transfer matrix}

\begin{figure}
\center
\includegraphics[scale=0.5]{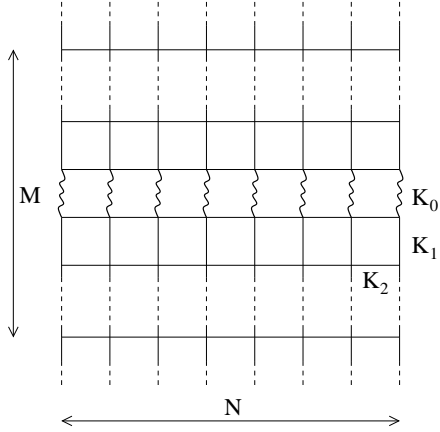}
\caption{Lattice and couplings in the strip geometry.}
\label{fig:latgeom}
\end{figure}

We now consider the strip and discretize the problem again by inserting a lattice
with $2M$ rows and $N$ columns. The couplings are isotropic, but we leave them
anisotropic for the moment. The basic unit is half the strip with $M$ rows and one
defect line in the middle. This is shown, for the case of a ladder defect, in Fig.
\ref{fig:latgeom}. The row transfer matrix for the whole unit will be called $W$. It has
the form $W=V^{M/2}V_{01}V^{M/2}$ where $V=V_1^{1/2}V_2V_1^{1/2}$ is the symmetrized row
transfer matrix for one step and $V_{01}$  describes the ladder. The quantities $V_1$ and 
$V_2$ contain the vertical and the horizontal bonds, respectively, and are given by
\eq{
\begin{split}
V_1 &= \ee^{K_1^* \sum_n \sigma_n^x}=\ee^{K_1^*\sum_n (2c_n^{\dag}c_n-1)} \\
V_2 &= \ee^{K_2 \sum_n \sigma_n^z \sigma_{n+1}^z} =
\ee^{K_2 \sum_n (c_n^{\dag}-c_n)(c_{n+1}^{\dag}+c_{n+1})}
\end{split}
\label{eq:v12}}
where we left out a prefactor in $V_1$ which is not important here. The second, 
fermionic forms arise after a spin rotation and a Jordan-Wigner transformation. The 
ladder operator is $V_{01}=V_{0}V_{1}^{-1}$ where $V_0$ has the same structure as $V_1$ 
with $K_0^*$ replacing $K_1^*$.

%
%
%
%

\begin{figure}
\center
\includegraphics[scale=0.7]{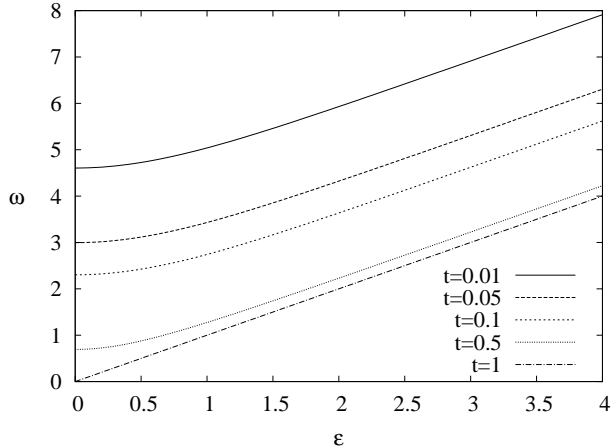}
\caption{Dispersion relation for the single-particle excitations in $W$ for several
values of the defect strength.}
\label{fig:disp}
\end{figure}

Although such product matrices as $V$ and $W$ can be diagonalized for the open 
boundaries we want to treat \cite{Abraham71}, the procedure is much simpler for 
periodic boundary conditions. We therefore assume these for the moment, which will 
give us the functional form of the eigenvalues. A Fourier transformation   
$c_n=\ee^{-i\pi/4} \frac{1}{\sqrt{N}}\sum_q \ee^{iqn}c_q$ then gives
\eq{
\begin{split}
& V_{1}^{1/2} = \prod_{q>0} \ex \left\{ K_1^* (c_q^{\dag}c_q+c_{-q}^{\dag}c_{-q}) \right\} \\
& V_{2} =\prod_{q>0} \ex \left\{ 2K_2 \left[ \cos q (c_q^{\dag}c_q+c_{-q}^{\dag}c_{-q})
- \sin q (c_q^{\dag}c_{-q}^\dag + c_{-q}c_{q}) \right]\right\}
\end{split}
\label{eq:v12q}}
where the momenta $q$ are different for even or odd total fermion number. The
problem now separates in $(q,-q)$ subspaces, and is solved by introducing new
operators $\alpha_q,\beta_q$ via a Bogoljubov transformation 
\cite{Schultz/Mattis/Lieb64}
\eq{
\left( \begin{array}{c}c_q \\ c_{-q}^{\dag}\end{array}\right) = 
\left(\begin{array}{cc}
\cos \varphi_q & -\sin \varphi_q \\
\sin \varphi_q & \cos \varphi_q
\end{array}\right)
\left( \begin{array}{c}\alpha_q \\ \beta_{q}^{\dag}\end{array}\right)
\label{eq:cantr}}
The diagonal form of the operator $V$ then reads, after changing to $k=\pi-q$
\eq{
V = A \, \ex \left(\sum_{k>0}\gamma_k (\alpha_k^\dag \alpha_k +\beta_k^\dag \beta_k )\right)
\label{eq:vdiag}}
with the well-known dispersion of the homogeneous system
\eq{
\ch \gamma_k = \ch 2K_1^* \,\ch 2K_2 - \sh 2K_1^* \,\sh 2K_2 \cos k
\label{eq:gammak}}
In a second step, one uses this result to diagonalize $W$. Some details are given
in Appendix A. $W$ has the same exponential form (\ref{eq:vdiag}) as $V$ but with other
operators and single-particle eigenvalues $\omega_k$ given by  
\eq{
\ch \omega_k = \ch 2\Delta \,\ch M\gamma_k + \sh 2\Delta \,\sh M\gamma_k \cos 2\varphi_k
\label{eq:omegak1}}
where $\Delta=K_0^*-K_1^*$. This is completely general. We now specialize to an isotropic 
critical system and small $k$. Then one has $\gamma_k=k$, 
$\cos 2\varphi_k = k/\sqrt{2}$ and the second term in (\ref{eq:omegak1}) can be
neglected. If one writes $Mk=\varepsilon_k$ for the eigenvalues of the homogeneous system, 
the relation becomes
\eq{
\ch \omega_k = \ch 2 \Delta \,\ch \varepsilon_k
\label{eq:omegak2}}
The basic feature is that the defect introduces a gap $2\Delta$ into the spectrum,
because it makes the system locally non-critical. The parameter $\ch 2\Delta$ can
be rewritten as
\eq{
\ch 2 \Delta = \frac {1}{2} \left(\frac {\thf K_0}{\thf K_1}
               +\frac {\thf K_1}{\thf K_0}\right) 
\label{eq:chdelta1}}
which gives via (\ref{tbond}) a very simple relation with the chain parameter $t$  
\eq{
\ch 2 \Delta = \frac {1}{2}\left (t+\frac {1}{t}\right)   
\label{eq:chdelta2}}
In Fig. \ref{fig:disp} we show the resulting dispersion curves for several values of $t$.
One sees that they become rapidly linear for larger values of $\varepsilon$ and 
show an upward shift, which from (\ref{eq:omegak2}) is $\ln\ch 2 \Delta$ and thus
varies logarithmically with $t$ for small $t$. These are exactly the features found 
in Fig. \ref{fig:spectra} for the numerical eigenvalues of the operator
$\mathcal{H}$. According to the remark at the end of section 3, these should correspond
to $2\omega_k$, since one needs two transfer matrices to build the full system. A
comparison of the numerical values of $2\omega(t)$ and $2\omega(t=1)$ verifies
(\ref{eq:omegak2}) relatively well, but there are shifts which may have to do with the open
boundaries. The allowed values of the momenta $k$ in this case have to be determined
from an equation which is known in the homogeneous case \cite{Abraham71} and will
be modified by the defect. In particular, they will vary with the defect strength.
%
%
%
%
\par
So far our consideration has been for a ladder defect in the lattice. The case of 
a chain defect can be obtained very simply by making a dual transformation in the
transfer matrices $V_1$ and $V_2$. This leaves the spectrum invariant, but changes 
vertical bonds into horizontal ones. The resulting chain defect then has coupling
$K_0^*$. Changing this into the desired $K_0$, one finds that $\ch 2 \Delta$ has
again the form (\ref{eq:chdelta2}) with $t$ now given by (\ref{tfield}). Thus bond and
field defects in the TI chain lead to the same transfer matrix spectra. 
\par
One should mention that a similar situation was found for CTM spectra in critical 
Ising models with a radial perturbation of the couplings decaying as $1/r$ 
\cite{Peschel/Wunderling92,Igloi/Peschel/Turban93}. This problem can be mapped to 
a homogeneous strip which is slightly non-critical. The dispersion relation then 
assumes the relativistic form $\omega_k=\sqrt{\Delta^2+\varepsilon_k^2}$. Our result 
reduces to this form for small $\Delta$.
\par
For the following considerations we define the parameter
\eq{
s = \frac {1}{\ch 2 \Delta} = \sin(2\arctan t)= \sin \left(2\arctan \frac{1}{t} \right)    
\label{eq:defs}}
which will turn out to be the quantity used in Ref. \cite{Sakai/Satoh08}.

\section{Entanglement entropy}

We are now in the position to evaluate $S$. For this we insert $2\omega_k$ as 
single-particle eigenvalues in the expression (\ref{ent-thermo}) and assume a large
value of $N$, so that the sum can be converted into an integral via
\eq{
\sum_{k} \rightarrow \frac {N}{\pi}\int_{0}^{\infty} \dd k =
     \frac {N}{M\pi}\int_{0}^{\infty} \dd \varepsilon
\label{eq:integration}}
The gradual shift of the allowed values of $k$ for free boundaries does not play a 
r\^ole in this limit. The mapping gives $2M/N=2\pi/\ln L$, so the prefactor of the 
second integral is $\ln L/\pi^2$ and we have the formula
\eq{
S = \frac{1}{\pi^2}I\ln L
\label{eq:entint}}
with 
\eq{
I = I_1 + I_2 =
\int_{0}^{\infty} \dd \varepsilon \; \ln(1+ \ee^{-2\omega}) +
\int_{0}^{\infty} \dd \varepsilon \;\frac {2\omega}{\ee^{2\omega}+1}
\label{eq:integral}}
Thus the entanglement entropy varies logarithmically and the coefficient is given
by the integrals in $I$. Inserting $\ch \omega = \ch \varepsilon /s$, it depends
on the single parameter $s$ which in turn is determined by the defect strength $t$.
In principle, this is the complete solution of the problem. 

\begin{figure}
\center
\psfrag{I(s)}[][][.8]{$\frac{12}{\pi^2}I(s)$}
\includegraphics[scale=0.7]{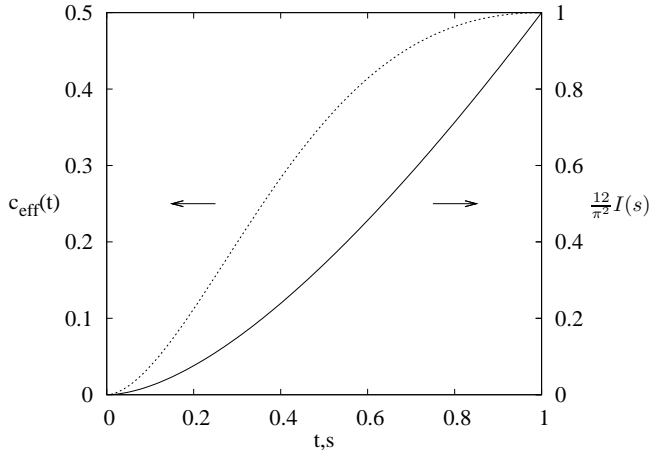}
\caption{Effective central charge $c_\mathrm{eff}(t)$ as a function of the defect 
strength $t$ and the normalized quantity $12\, I(s)/\pi^2$.}
\label{fig:ints}
\end{figure}

In the homogeneous case where $t=s=1$, the two integrals both have the value 
$\pi^2/24$ and one obtains the standard result $S=c/6 \ln L$ with $c=1/2$. For
general values of $s$, the integrals cannot be evaluated, but only brought into
a simpler form. Differentiating twice with respect to $s$ and substituting 
$x=\ch \varepsilon /s$ gives
\eq{
I''(s)=\frac{1}{s^2}\int_{1/s}^{\infty} \frac{\dd  x}{\sqrt{x^2-s^{-2}}}
\left[ \frac{x}{\sqrt{x^2-1}^3}\,\mathrm{arch\,}x - \frac{1}{x^2-1} \right]
\label{eq:int2diff}}
This can be evaluated using partial integrations with the simple result
\eq{
I''(s)= -\frac{1}{1-s^2}\ln s
\label{eq:int3diff}}
Integrating this again with the boundary values $I(0)=I(0)'=0$ then leads to
\eq{
I(s)=-\frac{1}{2} \bigg\{ \Big[ (1+s) \ln (1+s) + (1-s) \ln (1-s) \Big] \ln s
+ (1+s) \mathrm{Li_2}(-s) + (1-s) \mathrm{Li_2}(s) \bigg\}
\label{eq:intint}}
where $\mathrm{Li_2}(s)$ denotes the dilogarithm function defined by
\eq{
\mathrm{Li_2}(z)= - \int_{0}^{z} \dd x \frac {\ln(1-x)}{x}
\label{eq:dilog}}
The function $12/\pi^2I(s)$ is shown in Fig. \ref{fig:ints} and rises smoothly from 0 to 1
as $s$ increases. At $s=0$, it is non-analytic and varies as $(s^2/2)\ln(1/s)$. The 
quantity $c_\mathrm{eff}=6/\pi^2 I$ as a function of the defect strength $t$
is also shown in Fig. \ref{fig:ints} and varies between 0 and 1/2. It has the same 
non-analyticity
near $t=0$, found already in \cite{Igloi/Szatmari/Lin09}, but varies quadratically
near $t=1$. If one takes the numerical data of \cite{Igloi/Szatmari/Lin09}, they lie
perfectly on the curve. The same holds for the data on the XX model. Since $s(t)=s(1/t)$,
one obtains the same $c_\mathrm{eff}$ for weakened or strengthened defect bonds. This
was also observed in the numerics. According to the previous findings, field defects 
will also lead to the same results.
%
%
%
%
\par
\begin{figure}
\center
\psfrag{I(s)}[][][.8]{$\frac{12}{\pi^2}I_1(s)$}
\includegraphics[scale=0.7]{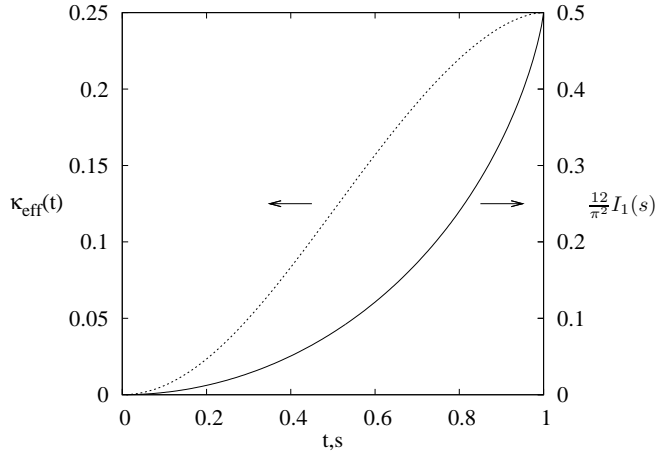}
\caption{Coefficient of the single-copy entanglement $\kappa_\mathrm{eff}(t)$ as a function of
the defect strength $t$ and the normalized quantity $12\, I_1(s)/\pi^2$.}
\label{fig:intsce}
\end{figure}

A quantity closely related to $S$ is the largest eigenvalue $w_1$ of the
reduced density matrix $\rho$. It plays a role in the so-called single-copy
entanglement problem \cite{Eisert/Cramer05}. If one writes $S_1 = -\ln w_1$,
the quantity $S_1$ is given by the first term in $I$
\eq{
S_1 = \frac{1}{\pi^2}I_1\ln L = \frac {\kappa_\mathrm{eff}}{6}\ln L
\label{eq:s1int}}
with a coefficient $\kappa_\mathrm{eff}$ analogous to $c_\mathrm{eff}$. Therefore 
$w_1$ varies as a power of $L$ and the exponent is $-\kappa_\mathrm{eff}/6$.
In the homogeneous system, $\kappa = c/2$  because then the integrals $I_1$ and 
$I_2$ have the same value. This is a general result for (homogeneous) conformally 
invariant systems \cite{Peschel/Zhao05,Orus06,Zhou06}. For arbitrary $s$, one can 
calculate $I_1$ in the same way as $I$. For the second derivative one finds 
\eq{
I_1''(s)= \frac{1}{1-s^2}+\frac{1}{2 s^2}\,\ln(1-s^2)
\label{eq:s1diff}}
which upon integration gives the simple result
\eq{
I_1(s)= \frac{1}{2} \Big[ \mathrm{Li}_2(s)+\mathrm{Li}_2(-s) \Big]
\label{eq:s1res}}
The quantity $12/\pi^2I_1(s)$ is shown in Fig. \ref{fig:intsce} and seen to rise from zero to
1/2 with a vertical slope at $s=1$. In $\kappa_\mathrm{eff}(t)$, the slope is zero, 
but the function is non-analytic, varying as $(1-t)^2\ln(1/(1-t))$ near $t=1$.
For small $t$, the behaviour is $6t^2/\pi^2$, which was found already in 
\cite{Igloi/Szatmari/Lin09} via perturbation theory. In \cite{Igloi/Szatmari/Lin09} 
the whole function $\kappa_\mathrm{eff}(t)$ was also calculated numerically and again 
the data match the formula given here.
%
%
%
%
\par
Finally, we want to make contact with the result found by Sakai and Satoh in 
\cite{Sakai/Satoh08}. There, in Eq. (4.6), a $c_\mathrm{eff}$ was given which 
consists of a term linear in a parameter $s$ $\emph{minus}$ an integral of bosonic 
origin which seems very different from the fermionic $I(s)$. However, its 
second derivative is exactly (\ref{eq:int3diff}) and it actually equals $I(s)$ 
as one can also verify by numerical integration. But it enters with a negative 
sign, and the additional linear term gives a finite slope at $s=0$. Therefore
the given expression cannot apply to our situation.

\section{Discussion}

Our calculations have given the asymptotic behaviour of the entanglement entropy 
between two halves of TI and XX chains separated by interface defects. In particular,
the effective central charge was found directly in terms of the corresponding defect 
parameter $t$. The same holds for $\kappa_\mathrm{eff}$ related to the largest 
density-matrix eigenvalue. The continuous variation of both quantities is analogous 
to that of the magnetic exponent $\beta_1$, and this analogy becomes even closer 
if one determines $\beta_1$ as in Appendix B. Thus all of them reflect the marginal 
character of the perturbation. 

Physically, one expects that these features are correlated with more elementary 
properties of the problem. 
For the XX chain, the obvious one is the transmission through the defect. This can be 
expressed through the two scattering phases one obtains from the S matrix. At the
Fermi surface for half filling ($k_F=\pi/2$), these are given by
\eq{
\delta_{\pm} = \pm \left(2 \arctan t -\frac{\pi}{2}\right)
\label{eq:scattering}}
and were already used in \cite{Peschel05} to guess an approximate expression for
 $c_\mathrm{eff}$. The transmission coefficient then is
\eq{
\mathcal{T}=\cos^2 \left(\frac {\delta_{+}-\delta_{-}}{2}\right) = \sin^2(2 \arctan t)=s^2
\label{eq:transmission}}
which gives a direct physical interpretation to the parameter $s$. The TI model is
not as straightforward, but one finds the same result. Such formulae also appear 
in the treatment of conformal defects \cite{Quella/Runkel/Watts07}.
They open the possibility to apply the results found here also to other situations.
The simplest case is the XX chain with arbitrary filling. Then it can be checked
numerically that the corresponding $\mathcal{T}(k_F)$ gives the correct $c_\mathrm{eff}$. But 
one can also consider more complicated types of defects, where a direct calculation 
as done here would be quite difficult \cite{Eisler10}.
An interesting problem would also be the extension to a quench from a homogeneous 
system to one with a defect. The numerics show, that $c_\mathrm{eff}$ then appears
in the time dependence \cite{Igloi/Szatmari/Lin09}.

\begin{acknowledgments}
We thank Ferenc Igl\'oi for correspondence and for making his data available to us
and Malte Henkel for discussions. I.P. also thanks the Niels Bohr Institute for its 
hospitality at the final stage of this work. V.E. acknowledges financial support by 
the Danish Research Council, QUANTOP and the EU projects COQUIT and QUEVADIS.
\end{acknowledgments}

\appendix

\section*{APPENDIX A.}

In this appendix, we give some more details on the diagonalization of $W$ for
completeness. Such calculations have, of course, been done also in other studies dealing 
with layered Ising lattices, see e.g. \cite{Becker/Hahn72, Au-Yang/McCoy74,
Hoever/Wolff/Zittartz81}. 

We begin with $V$. Forming Heisenberg operators with $V_1$ and $V_2$, one has 
\eq{
\begin{split}
V_1^{1/2} &\left( \begin{array}{c}c_q \\ c_{-q}^{\dag}\end{array}\right) V_1^{-1/2} =
\left(\begin{array}{cc} \ee^{-K_1^*} & 0 \\ 0 & \ee^{K_1^*} \end{array}\right)
\left( \begin{array}{c}c_q \\ c_{-q}^{\dag}\end{array}\right) \\
V_2 &\left( \begin{array}{c}c_q \\ c_{-q}^{\dag}\end{array}\right) V_2^{-1} =
\left(\begin{array}{cc} C_2-S_2 \cos q & S_2 \sin q \\ S_2 \sin q &  C_2+S_2 \cos q 
\end{array}\right)
\left( \begin{array}{c}c_q \\ c_{-q}^{\dag}\end{array}\right)
\end{split}
\label{eq:v12heis}}
with the notation $C_i=\ch 2K_i$ and $S_i=\sh 2K_i$ for $i=1,2$. This gives for $V$
\eq{
V \left( \begin{array}{c}c_q \\ c_{-q}^{\dag}\end{array}\right) V^{-1} =
\left(\begin{array}{cc} 
\ee^{-2K_1^*} (C_2-S_2 \cos q) & S_2 \sin q \\
S_2 \sin q &  \ee^{2K_1^*} (C_2+S_2 \cos q)
\end{array}\right)
\left( \begin{array}{c}c_q \\ c_{-q}^{\dag}\end{array}\right)
\label{eq:vheis}}
The matrix on the right has determinant 1 and thus eigenvalues $\ee^{\pm\gamma_q}$ where
\eq{
\ch \gamma_q = C_1^* C_2 + S_1^* S_2 \cos q
\label{eq:gammaq}}
It is diagonalized by the canonical transformation (\ref{eq:cantr}) where
the angle is, for the isotropic system at the critical point
\eq{
\cos 2\varphi_q = \sqrt{\frac{2\cos^2 \frac{q}{2}}{1+\cos^2 \frac{q}{2}}}
\label{eq:cos2phi}}
The matrix $V$ then has the diagonal form given in (\ref{eq:vdiag}). 

In the second step one needs the Heisenberg operators of $\alpha_q$ and $\beta_{q}$ 
with $V^{M/2}$ and $V_{01}$. These are
\eq{
\begin{split}
V^{M/2} &\left( \begin{array}{c} \alpha_q \\ \beta_{q}^{\dag}\end{array}\right) V^{-M/2} =
\left(\begin{array}{cc} \ee^{-M\gamma_q/2} & 0 \\ 0 & \ee^{M\gamma_q/2} \end{array}\right)
\left( \begin{array}{c}\alpha_q \\ \beta_{q}^{\dag}\end{array}\right) \\
V_{01} &\left( \begin{array}{c} \alpha_q \\ \beta_{q}^{\dag}\end{array}\right) V_{01}^{-1} =
\left(\begin{array}{cc} c^2 \ee^{-2\Delta}+s^2 \ee^{2\Delta} & cs(\ee^{2\Delta}-\ee^{-2\Delta}) \\
cs(\ee^{2\Delta}-\ee^{-2\Delta}) & c^2 \ee^{2\Delta}+s^2 \ee^{-2\Delta}
\end{array}\right)
\left( \begin{array}{c} \alpha_q \\ \beta_{q}^{\dag}\end{array}\right)
\end{split}
\label{eq:vm01}}
where $\Delta=K_0^*-K_1^*$, $c=\cos \varphi_q$ and $s=\sin \varphi_q$ . This finally gives
for the transfer matrix $W$
\eq{
W \left( \begin{array}{c} \alpha_q \\ \beta_{q}^{\dag}\end{array} \right) W^{-1} =
V^{M/2}V_{01}V^{M/2}
\left( \begin{array}{c} \alpha_q \\ \beta_{q}^{\dag}\end{array} \right)
V^{-M/2}V_{01}^{-1} V^{-M/2}=
\left( \begin{array}{cc} a & d \\ d & b \end{array} \right)
\left( \begin{array}{c} \alpha_q \\ \beta_{q}^{\dag}\end{array} \right)
\label{eq:wheis}}
with
\eq{
\begin{split} 
a &= (\ch 2\Delta -\sh 2\Delta \cos 2\varphi_q)\ee^{- M\gamma_q} \\
b &= (\ch 2\Delta +\sh 2\Delta \cos 2\varphi_q)\ee^{M\gamma_q} \\
d &= \sh 2\Delta \sin 2\varphi_q.
\end{split}
\label{eq:coeff}}
Again the determinant equals 1 and the eigenvalues have the form $\ee^{\pm\omega_q}$. 
This leads to the result (\ref{eq:omegak1}) in the text.

\section*{APPENDIX  B.}

The local magnetic exponent $\beta_1$ was used in section 3 to find the correct
corespondence between the defect parameters in the chain and in the Ising lattice.
If $\beta_1$ were not known, one could also have calculated it with our approach. In
the CTM method, order parameters are obtained by 
fixing the spins at the outer boundary and then calculating the expectation 
value $m_0$ of the central spin. For free fermionic systems, this gives $m_0$ 
as a product involving the single-particle CTM eigenvalues \cite{Igloi/Peschel/Turban93}. 
Taking the logarithm and converting the sum into an integral via (\ref{eq:integration})
one finds in our notation 
\eq{
\ln m_0 = \ln L \,\frac {1}{\pi^2}\int_{0}^{\infty} \dd \varepsilon \ln \thf \omega
\label{eq:betaint}}
The coefficient of $\ln L$ is then $-\beta_1$. The same formula was used also in 
\cite{Peschel/Wunderling92}. In contrast to $I(s)$, this integral can be evaluated
in closed form. Substituting $z= 1/\ch \varepsilon$ one has
\eq{
\beta_1 = -\frac{1}{2\pi^2}\int_{0}^{1}\frac{\dd z}{z\sqrt{1-z^2}}
\left[ \ln (1+sz) + \ln (1-sz) \right] = \frac{1}{2\pi^2}\arcsin^2 s
\label{eq:betaint2}}
Since $s= \sin (2 \arctan t) = \sin (2 \arctan \frac{1}{t})$, one has to
make a choice here. Taking the second form gives 
\eq{
\beta_1 = \frac{2}{\pi^2} \arctan^2 \left(\frac{1}{t}\right) 
\label{eq:beta}}
which is the ladder result (\ref{betaladder}) if one writes it in terms of the lattice 
parameters. However, since (\ref{eq:betaint2}) cannot exceed the value $\beta_1=1/8$ of
the homogeneous system, this expression is valid only for $ t\ge 1$, or $K_0 \ge K_1$.
For $t<1$, one has to take the first form for $s$. But in this case, when the system is
slightly in the disordered phase, the fixed boundaries lead to an additional low-lying
eigenvalue $\omega_0$ wich vanishes as a power of $L$ and thus contributes to $\beta_1$.
Such a situation was already found in \cite{Peschel/Wunderling92}, and through this
mechanism (\ref{eq:beta}) will be obtained also for $t<1$.
The case of the chain defect is analogous.

\pagebreak

\end{document}